\documentclass[a4paper]{jpconf}
\usepackage{hyperref}
\usepackage{amsmath}
\usepackage{amssymb}
\usepackage{slashed}
\usepackage{graphicx}
\usepackage{graphics}
\usepackage{epsfig}
\usepackage{color}
\newcommand{\Eqref}[1]{Eq.~\eqref{#1}}

\begin{document}
\title{Observable consequences of quantum gravity: Can light fermions exist?}
\author{A. Eichhorn \footnote{astrid.eichhorn@uni-jena.de}}
\address{Theoretisch-Physikalisches Institut, Friedrich-Schiller-Universit{\"a}t Jena, Max-Wien-Platz 1, D-07743 Jena, Germany}
\begin{abstract}
Any theory of quantum gravity must ultimately be connected to observations. This demand is difficult to be met due to the high energies at which we expect the quantum nature of gravity to become manifest. Here we study, how viable quantum gravity proposals can be restricted by investigating the interplay of gravitational and matter degrees of freedom. Specifically we demand that a valid quantum theory of gravity must allow for the existence of light (compared to the Planck scale) fermions, since we observe these in our universe. Within the effective theory framework, we can thus show that UV completions for gravity are restricted, regardless of the details of the microscopic theory. Specialising to asymptotically safe quantum gravity, we find indications that universes with light fermions are favoured within this UV completion for gravity. 
\end{abstract}
\section{How to connect quantum gravity to observations}
In quantum gravity, the goal is the construction of an internally consistent theory, which can then be tested against experiment. Since the typical scale of quantum gravity effects is generally expected to be $\mathcal{O}(M_{\rm Planck})$, devising doable experiments that test the quantum nature of gravity is highly challenging. 
Hence it is considerably simpler to \emph{indirectly} test quantum gravity: Clearly the observed low-energy properties of matter must be compatible with a quantum theory of gravity, if it is to describe our universe. A particular quantity that can be sensitive to quantum gravity fluctuations is constituted by fermion masses. Within the Standard Model, these arise from chiral symmetry breaking ($\chi$SB), either induced by the strong interactions, or by the Higgs sector. In a similar fashion, quantum gravity fluctuations might induce strong correlations in the fermion sector and lead to $\chi$SB. In this case, the induced fermion masses would naturally be $\mathcal{O}(M_{\rm Planck})$, analogously to QCD, where the fermion masses induced by $\chi$SB are comparable to $\Lambda_{\rm QCD}$. Since our universe contains fermions which are considerably lighter than the Planck scale, we conclude that any viable theory of quantum gravity must evade such a mechanism.

Here, we will show that this requirement allows to put restrictions on quantum gravity theories. 
We first investigate generic UV completions for gravity by making use of the framework of effective field theories: Irrespective of the UV completion for gravity, an effective parameterisation of quantum gravity fluctuations in terms of metric fluctuations should hold on scales up to a UV scale $\Lambda \lesssim M_{\rm Planck}$. (This allows, e.g. to compute quantum corrections to the Newtonian potential, see, e.g. \cite{Donoghue:1993eb}.) Then the UV completion for gravity determines the values of the couplings in the effective theory on the UV cutoff scale $\Lambda$. A Renormalisation Group (RG) study allows to investigate, whether the predicted values for the couplings are compatible with the existence of light fermions on lower scales.

A parameterisation of quantum gravity fluctuations in terms of metric fluctuations is also central to a specific UV completion for gravity, namely the asymptotic-safety scenario \cite{Weinberg:1980gg}, where the gravitational degrees of freedom are carried by the metric up to arbitrarily high momentum scales. The far UV is then dominated by an interacting fixed point in the running couplings. Evidence for the existence of this fixed point has been collected, e.g. in \cite{AS}; for reviews see \cite{AS_reviews}. 

The main difference between these two settings is, that in the latter case the requirement of an interacting fixed point determines the UV behaviour at arbitrarily high momenta. In the former setting, our analysis extends over a finite range of scales, only. The values of the couplings at the finite UV-scale $\Lambda$ are then not restricted by a fixed-point requirement, and are determined by the microscopic theory.

\section{How to investigate the existence of light fermions}
Within the framework of the functional RG, $\chi$SB is signalled by divergent four-fermion couplings. This connection arises as follows: Introducing composite bosonic degrees of freedom, schematically $\phi \sim \bar{\psi} \psi$, in the path integral, allows to rewrite a four-fermion coupling $\lambda \left(\bar{\psi}\psi\right)^2 \sim \phi \bar{\psi} \psi + \frac{1}{\lambda}\phi^2$. Within the bosonic picture, the onset of spontaneous symmetry breaking is signalled by the bosonic mass going to zero. Thus $\lambda \rightarrow \infty$ signals the onset of spontaneous $\chi$SB in the purely fermionic language, which we will use here\footnote{The introduction of composite bosonic degrees of freedom becomes necessary only for an investigation of the symmetry-broken regime. A purely fermionic formulation suffices to detect the onset of symmetry breaking, which is all we will be interested in here.}.
 
To investigate, whether the four-fermion couplings diverge, we study their $\beta$ functions. 
We use a functional RG equation, the Wetterich-equation \cite{Wetterich:1993yh} for the scale-dependent effective action $\Gamma_k$, which is the generating functional of 1PI correlators
that include all fluctuations from the UV down to the infrared (IR) scale $k$. At $k=0$, $\Gamma_k$ coincides with the standard effective
action $\Gamma=\Gamma_{k=0}$. The scale derivative of $\Gamma_k$ is given by 
\begin{equation}
 \partial_t \Gamma_k = \frac{1}{2}{\rm STr} \{
 [\Gamma_k^{(2)}+R_k]^{-1}(\partial_t R_k)\}. \label{floweq}
\end{equation}
Here, $\partial_t = k \,\partial_k$; $\Gamma_k^{(2)}$ is the second functional
derivative of $\Gamma_k$ with respect to the fields, and $R_k$ is an IR
regulator function. 
The supertrace $\rm STr$ contains a trace over the spectrum of the full
propagator (for reviews, see
\cite{reviews}). This equation is exact, and is applicable in perturbative as well as non-perturbative settings, since its derivation does \emph{not} rely on the existence of a small parameter. 
Practical calculations do however usually require truncations of all possible operators in the effective action to a, typically finite, subset.

In choosing an appropriate truncation it is crucial that we include all four-fermion couplings compatible with the symmetries, since any of these might diverge and induce $\chi$SB. We thus study a Fierz-complete basis of ${\rm SU}({\rm N}_f)_L \times {\rm SU}({\rm N}_f)_R$ symmetric four-fermion interactions, where ${\rm N}_f$ is the number of fermions, and $i,j=1,...,{\rm N}_f$. 
\begin{eqnarray}
 \Gamma_{k\, \rm F}&=& \frac{1}{2} \int d^4x \, \sqrt{g}\,
 \bigl[\bar\lambda_-(k) (V-A) + \bar\lambda_+(k)
(V+A ) \bigr],\\ \nonumber
\mbox{where }  V&=& \left( \bar{\psi}^i \gamma_{\mu}\psi^i \right)\left( \bar{\psi}^j
   \gamma^{\mu}\psi^j \right), \quad \quad A=- \left( \bar{\psi}^i \gamma_{\mu}\gamma^5\psi^i \right)\left(
  \bar{\psi}^j \gamma^{\mu}\gamma^5\psi^j \right). 
\end{eqnarray}

Generically, the $\beta$ functions of the dimensionless renormalised couplings $\lambda_\pm = \frac{k^2 \bar\lambda_\pm}{Z_\psi}$ read \cite{Eichhorn:2011pc}
\begin{equation}
 \beta_{\lambda_{\pm}}= (2+\eta_{\psi})\lambda_{\pm}+a \,\lambda_{\pm}^2 +b\, 
\lambda_{\pm}\lambda_{\mp} + c \,\lambda_{\mp}^2+d \lambda_{\pm}+e,
\label{eq:betalambda}
\end{equation}
where $\eta_{\psi} = - \partial_t \ln Z_{\psi}$ is the fermionic anomalous dimension related to the wave-function renormalisation, see below.
Herein the first term arises from dimensional (and anomalous) scaling and reflects the perturbative non-renormalisability of four-fermion couplings in four dimensions. The
quadratic contributions follow from a purely fermionic two-vertex diagram. A tadpole contribution $\sim
d\lambda_{\pm}$ may also exist, as well as a $\lambda_{\pm}$-independent part
$\sim e$ resulting from the coupling to other fields, e.g. to the metric, thus being decisive for the question of $\chi$SB.

\begin{figure}[!here]
\begin{minipage}{0.28\linewidth}
\includegraphics[scale=0.52]{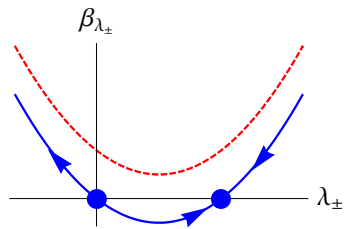}
\end{minipage}
\begin{minipage}{0.7\linewidth}
\caption{In the absence of further interactions, the parabola-type
  $\beta$ function (solid blue line) exhibits two fixed points, the Gau\ss ian one at
  $\lambda_{\pm}^\ast=0$ and a non-Gau\ss ian one at
  $\lambda_{\pm}^\ast=\lambda_{\pm, \text{cr}}>0$. Arrows indicate the RG
  flow towards the IR, and initial conditions to the left of $\lambda_{\pm}=0$ lead to $\chi$SB. Further interactions can shift the parabola (red dashed
  line) such that no fixed points exist. Such a scenario typically occurs in QCD-like theories \cite{Kondo:1991yk,Gies:2003dp,Gies:2005as}.} 
\label{fig:parabolasketch}
\end{minipage}
\end{figure}

The crucial observation is that whenever fixed points exist, initial conditions for the RG flow can be found such that chiral symmetry breaking is avoided, see fig. \ref{fig:parabolasketch}. 

To evaluate the gravitational contribution to the $\beta$ functions, we include terms carrying dynamics for gravitational as well as fermionic degrees of freedom in the following truncation
\begin{eqnarray}
 \Gamma_k = \frac{Z_N(k)}{16 \pi G_N}\int 
d^4 x \sqrt{g}(-R+ 2 \bar{\lambda}(k))+ \Gamma_{k\,\rm gf}+ \int d^4x \sqrt{g}\, i Z_{\psi}
\bar{\psi}^i
\gamma^{\mu}\nabla_{\mu}\psi^i
+ \Gamma_{k\, \rm F},
\end{eqnarray}
Herein the
bare Newton constant $G_{\text{N}}$ is related to the dimensionless renormalised Newton coupling $G(k)= k^2 Z_N(k)G_N$ and we have included a cosmological constant $\bar\lambda(k)=k^2 \lambda(k) $. $\Gamma_{k\, \rm gf}$ denotes a standard gauge-fixing and ghost term in Landau deWitt background field gauge, see, e.g. \cite{Eichhorn:2010tb}. Our truncation is motivated by similar truncations in QCD, which allow for a determination of the critical temperature for $\chi$SB \cite{Braun:2005uj}. The $\beta$ functions for the Newton coupling and the cosmological term have been analysed in this truncation (with $Z_{\psi}=1$) in \cite{Percacci:2002ie}.

\section{Restricting UV completions for gravity}

The main question is whether quantum gravity fluctuations lead to fixed-point annihilation in the $\beta$ functions.
As our main result we observe that this is not the case. We find that for any $\{G>0, \lambda, N_f\}$ the system of $\beta$ functions $\beta_{\lambda_{\pm}}$ shows four fixed points (for details of the calculation, see \cite{Eichhorn:2011pc}). Thus metric fluctuations do not contribute strongly to $\lambda_{\pm}$-independent terms in \Eqref{eq:betalambda}, which could lead to fixed-point annihilation. Instead, we find that gravity fluctuations mainly enhance anomalous scaling terms $\sim \lambda_{\pm}$. Thus our first conclusion is that, although gravity is an attractive force, metric fluctuations do not automatically induce strong fermionic correlations and lead to $\chi$SB and bound state formation.

Thus $\chi$SB is a question of \emph{initial conditions} for the RG flow, cf. fig. \ref{fig:parabolasketch}. Here lies the main difference between the asymptotic-safety scenario and other UV completions for gravity: In the former, the system has to sit on a fixed point in the far UV. If there were no fixed points in $\beta_{\lambda_{\pm}}$, asymptotic safety could not constitute a UV completion for gravity with dynamical fermions. 
Here, we have four fixed points with different universal properties at our disposal, which all define a valid UV completion. The number of relevant couplings (connected to the free parameters of the theory) is either zero, one or two, depending on the fixed point. Thus these fixed points constitute UV completions with varying predictive power. Their existence entails that metric fluctuations do not directly induce $\chi$SB and the scenario is compatible with observations.

\begin{figure}[!here]
\begin{minipage}{0.45\linewidth}
  \includegraphics[scale=0.45]{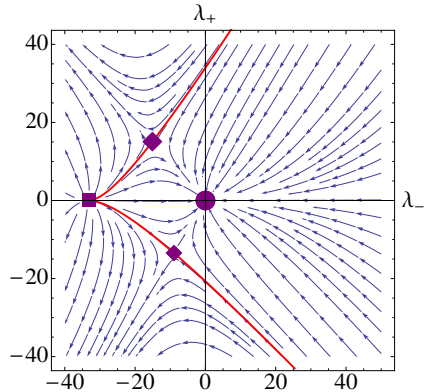}
\end{minipage}
\begin{minipage}{0.52\linewidth}
\caption{Flow towards the IR in the $\lambda_{+},
  \lambda_{-}$-plane for $\eta_N =0, \eta_{\psi}=0, G=0.1, \lambda=0.1$ and
  $N_f=6$. For initial values to the right of the red lines the chiral system
  is in the universality class of the (shifted) Gau\ss ian fixed point. Any
  microscopic theory that would put the effective quantum field theory to the
  left of the red lines would generically not support light fermions; thus the initial conditions are restricted to lie in the basin of attraction of the Gau\ss{}ian fixed point.}
\label{effplot}
\end{minipage}
\end{figure}

In the case of other UV completions for gravity, there is no fixed-point requirement, and in fact the initial conditions for the RG flow can lie anywhere in the space of all couplings. Here, the crucial observation is that unbroken chiral symmetry restricts the allowed initial conditions, see fig. \ref{effplot}. 
We thus conclude that the existence of light fermions puts restrictions on any UV completion for gravity. 

In conclusion, we have found - within a truncation of the full RG equations - that, although gravity is an attractive force, even strong metric fluctuations do not directly lead to $\chi$SB and bound state formation for fermions. This result applies to asymptotically safe quantum gravity, as well as other UV completions, where we show that the requirement of unbroken $\chi$SB can be used to restrict the space of couplings. 

\ack I thank Holger Gies for helpful comments on this manuscript. This work was supported
by the DFG-Research Training Group "Quantum- and
Gravitational Fields" (GRK 1523/1).

\end{document}